\documentclass[11pt,preprint]{aastex}

\usepackage{epstopdf}
\usepackage[normalem]{ulem}
\usepackage{graphicx}

\shorttitle{Star-forming Region Sh 2-233IR  I. Embedded Clusters}
\shortauthors{Yan et al.}

\begin{document}

\title{Star-forming Region Sh 2-233IR  I. Deep NIR Observations toward the Embedded Stellar Clusters}

\author{Chi-Hung Yan}
\affil{Institute of Astronomy and Astrophysics, Academia Sinica, \\
P.O. Box 23-141, Taipei 10617, Taiwan}
\affil{Department of Earth Sciences, National Taiwan Normal University, \\
Taipei 10617, Taiwan}
\email{chyan@asiaa.sinica.edu.tw}

\author{Y. C. Minh}
\affil{Korea Astronomy and Space Science Institute \\
 838 Daeduk-daero, Yuseong-gu, Daejeon 305-348, Korea}

\author{Shiang-Yu Wang \& Yu-Nang Su}   
\affil{Institute of Astronomy and Astrophysics, Academia Sinica \\
P.O. Box 23-141, Taipei 10617, Taiwan}

\author{Adam Ginsburg}   
\affil{Center for Astrophysics and Space Astronomy \\
University of Colorado, 389 UCB, Boulder, CO 80309-0389, USA}

\begin{abstract}
We observed the Sh~2-233IR (S233IR) region with better sensitivity in near-infrared than previous studies for this region.  By applying statistical subtraction of the background stars, we identified member sources and derived the age and mass of three distinguishable sub-groups in this region: Sh~2-233IR NE,  Sh~2-233IR SW, and the ``distributed stars'' over the whole cloud.  Star formation may be occurring sequentially with a relatively small age difference ($\sim0.2-0.3$ Myrs) between subclusters.   We found that the slopes for initial mass function ($\Gamma\sim -0.5$) of two subclusters are flatter than that of Salpeter, which suggests that more massive stars were preferentially formed in those clusters compared to other Galactic star-forming regions.  These subclusters may not result from the overall collapse of the whole cloud, but have formed by triggering before the previous star formation activities disturbed the natal molecular cloud.   Additionally, high star formation efficiency ($\gtrsim$40\%) of the subclusters may also suggest that stars form very efficiently in the center of NE.
\end{abstract}


\section{Introduction}
Within 2 kpc, $70-90$\% of stars have been found to form in the clusters \citep{lada03}.   These newly formed stars, however, are embedded deeply in dense molecular cloud and there exists various difficulties in identifying sources and deriving physical parameters.  Several statistical tools, such as the K-band luminosity function (KLF), the color-color or color-magnitude diagrams, etc., have been successfully used to constrain the characteristics of deeply embedded stellar clusters \citep[e.g.,][]{lada03}.   Development of new sensitive equipments, such as, large mosaic IR array, enables us to observe embedded stellar cluster in greater detail than previous effort.  The Sh~2-233IR region is a well studied star-forming region in our Galaxy \citep[e.g.,][]{porras00} and it provides an ideal laboratory for understanding properties of embedded clusters.  Therefore, we revisit this region with higher sensitivity near-infrared data taken toward a wider region to better constrain the properties of the embedded stellar cluster than previous studies.

Toward the direction of the Galactic Anticenter, Sh~2-233IR (hereafter S233IR), as a part of the Sh~2-235 GMC complex \citep{ry08}, is located at a distance of about 1.8 kpc \citep{porras00}, in association with four extended HII regions, Sh~2-231, 232, 233, and 235 \citep{heyer96}.  Its position coincides with an IRAS source, IRAS 05358+3543 ($\alpha_{2000}=05^h39^m10^s$ $\delta_{2000}=+35^{\circ}45^\prime19^{\prime\prime}$).  S233IR is classified as a massive star formation region \citep{sri02}, showing CO outflows \citep{snell90} and various maser emissions associated with this region  \citep{henning92,tofani95,beuther02b,menten91,minier05}.    The $K^\prime$ band image of this region shows many bright stellar sources and also extended nebulous features associated with dust emission \citep{hodapp94}.  The two embedded young clusters, Sh~2-233IR SW (hereafter SW, located in the south-west direction from the center) and Sh~2-233IR NE (hereafter NE, in the north-east direction), are notable in this region, with remarkable H$_2$ bow shocks associated with the NE cluster \citep{porras00}.  
In addition, numerous studies have been reported, especially for the NE cluster, including polarimetric observations \citep{jiang01,yao00},  molecular outflows \citep{beuther07,mz04,beuther02a,cesaroni99,lari99}, and mid-infrared sources \citep{longmore06}.  

In this {\it paper}, we revisit this S233IR region with wider field of view, higher resolution, and better sensitivity data in near-infrared and radio wavelengths.   Our goal is to understand star formation history in terms of age, star formation efficiency and initial mass function.  We summarize our observations at radio and near-infrared in \S~\ref{obs}, and the observed results in \S~\ref{results}. Discussion is given in \S~\ref{discuss}, and the summary in \S~\ref{summary}.  This is the first paper in the series of our work for the S233IR and associated region.  Here we focus on properties of the embedded stellar clusters in the S233IR region.  The H$_2$ shock features and cloud kinematics have been studied by \cite{adam09}, and the extended nebulous $K$-band features, young stellar object and star formation in a larger area will be discussed in the second paper \citep{chy09b}.

\section{Observations and Data Analysis}\label{obs}

\subsection{CFHT observations}
Near infrared data were obtained using the Wide-field Infrared Camera \citep[WIRCam,][]{pascal04} equipped on Canada-France-Hawaii Telescope\footnote{The Canada-France-Hawaii Telescope (CFHT) is operated by the National Research Council of Canada, the Institut National des Sciences de l'Univers of the Centre National de la Recherche Scientifique of France,  and the University of Hawaii.} (CFHT) on November 18, 19 and December 20, 2005 for the H$_2$ and $K_S$ bands,  and February 4, 2007 for the $J$, $H$, and $K$ continuum bands.  Its field of view is $20\arcmin\times20\arcmin$ with a pixel scale of 0.3$\arcsec$ per pixel, and  the seeing was $0.5\arcsec - 0.7\arcsec$ during the observations.   Exposure times on each frame were 6 and 30 (58 for H$_2$) seconds for the wide band and the narrow band filters, respectively.   Total integration times for each filter were 420 seconds for $J$ and $H$, 235 seconds for $K_S$, 1755 seconds for H$_2$, and 2400 seconds for Br$\gamma$ and $K$ continuum.  The CFHT WIRCam standard pipeline (the IDL Interpretor of the WIRCam Images, aka i'iwi) was used for basic image processing, such as, the standard bias subtraction, flat-fielding and bad pixel masking.  Sky background of each image was subtracted using the comoving-averaged frame taken during the observations.    Astrometry and photometry corrections were performed against the 2MASS catalog in order to correct the image distortion and flux level.   And then the images were analyzed using IDL, IDL Astronomy Library and PhotVis, a GUI interpretation of IDL DAOPHOT package.   The completeness and limiting magnitude were estimated by adding artificial stars in the observed images.   The limiting magnitudes were estimated to be 21.0, 20.0 and 19.0 for the $J$, $H$, and $K_S$ bands, respectively, at a 90\% completeness level.  The saturation level of each filter is 44000 ADUs, corresponding to 13.4 ($J$), 13.5 ($H$) and 12.8 ($K_S$) magnitude, and the photometry for saturated stars were adopted from the 2MASS catalog directly.    The error of photometry at faint end is $\sim0.6$ magnitude.  Within the observed region, we identified 567, 715 and 714 stars in the $J$, $H$, and $K_S$ bands, respectively.  

\subsection{ARO observations}
The CO line ($J=3-2$, 345.7959899 GHz) observations were carried out with the SMT\footnote{The Submillimeter Telescope is  operated by the Arizona Radio Observatory (ARO), Steward Observatory, University of Arizona, with partial support from the Research Corporation.} 10 m telescope at Mount Graham, Arizona, using the dual-channel SIS 0.8 mm receivers operated in single-sideband dual polarization mode with image sideband rejection of at least 16 dB on April 6, 8, 10, 25 and 28, 2008.  The spectrometers used were a 2048 channel acousto-optical spectrometer (AOS) with a spectral resolution of 500 kHz per channel and 1024 channel Forbes Filterbanks (FFBs) with a spectral resolution of 1 MHz per channel.   All spectrometers were used simultaneously.  On-the-fly (OTF) mapping, centered at $\alpha=05^h39^m10^s$ $\delta=+35^{\circ}45'19''$, was used to map a $20\arcmin\times20\arcmin$ area.   The total number of scanning rows was 120 with 10$\arcsec$ spacing and a scanning rate of 10$\arcsec$/sec.   The calibration was done using the absolute position switching mode after finishing every individual row.   System temperatures varied from 500 to 1500 K during observations depending on the weather conditions.  The observed data were calibrated to the antenna temperature scale, $T_{A}$, which was corrected for atmospheric attenuation,  and then converted to main-beam temperature by $T_{mb}=\frac{T_A}{\eta_{mb}}$, where the main-beam efficiency, $\eta_{mb}$, is $\sim$0.65 (see the ARO web-site, http://aro.as.arizona.edu/, for the details).

\section{Results}\label{results}

\subsection{The S233IR molecular cloud}\label{cloud}
\begin{figure}
\centering
\resizebox{15cm}{!}{\rotatebox{0.}{\includegraphics{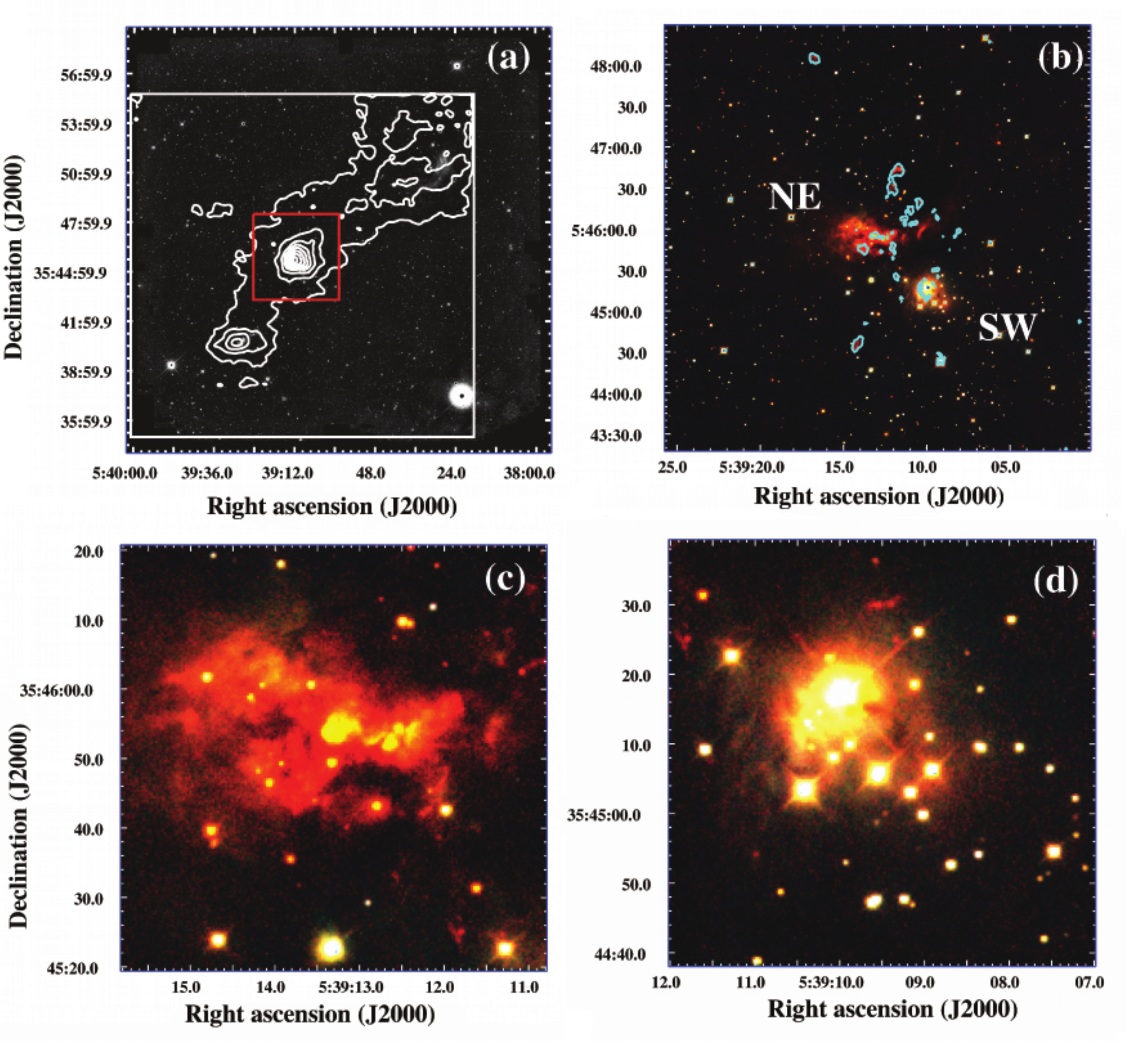}} }
\caption{(a) The CO $J=3-2$ integrated intensity map in the velocity range of $v_{lsr} = -30 - 0$ km s$^{-1}$ ({\it contours}), overlaid on the the WIRCam $K_S$ band image of the size $25\arcmin \times 25 \arcmin$.   The outer white box is the region observed with the SMT using the on-the-fly mapping mode and the inner red box represents the S233IR region.  Contour levels of the CO map were shown from 20 to  100 K km s$^{-1}$ with spacings of 10 K km s$^{-1}$.  The peak flux density is 177 K km s$^{-1}$ at the center.  (b) The color composite image near S233IR made using $J$ (blue), $H$ (green) and $K_S$ (red) images.  The H$_2$ emission is overlaid with contours on the composite color image.  The lowest H$_2$ contour is shown as flux level to be 0.27$\times10^{-3}$ mJy. The image center is at $\alpha=05^h39^m10^s$ $\delta=+35^{\circ}45'19''$ with the size of about 5$\arcmin$ by 5$\arcmin$.  (c) The color image of NE cluster made using $J$ (blue), $H$ (green) and $K_S$ (red) filters.  (d) The color image of SW cluster made using $J$ (blue), $H$ (green) and $K_S$ (red) filters.
}
\label{aro}
\end{figure}

 The CO $J=3-2$ emission shows that the molecular cloud associated with S233IR is located on the ridge connecting two other star forming regions in the northwest and southeast directions about 3.5 pc away from the S233IR cloud at a distance of 1.8 kpc (Fig.~\ref{aro}(a)).  The northwest region coincides with the HII region Sh~2-233 \citep[IRAS 05351+3549,][]{casoli86}, which is excited by a B1.5II star \citep{hunter90}.   High-velocity CO $J=1-0$ emission was found toward this region \citep{jiang00}.   Towards the southeast of S233IR, there is another star forming core called G173.58+2.45 \citep[IRAS 05361+3539,][]{sc96}, 
where a small YSO cluster \citep{varricatt05} and a molecular outflow exists in the east-west direction \citep{sw02}.

The main goal of the CO $J=3-2$ observations was to determine the boundary of the molecular cloud embedding the S233IR cluster.  Fig.~\ref{aro}(a) shows that the dense region of the S233IR cloud is centered on the density peak with a roundish shape.  The two extended $K_S$ emission features near the center of the CO $J=3-2$ emission correspond to the NE and SW clusters.   At the 20 K km s$^{-1}$ level, there is a common envelope for three star formation regions.  The average size of the 30 K km s$^{-1}$ contour level of the CO ($J=3-2$) emission is about 1.5 pc and the half-power width (HPW) is $\sim$0.8 pc.  We set the boundary of the S233IR cloud with this 30 K km s$^{-1}$ ($r\sim$2.5\arcmin) contour level and checked the member sources of the embedded cluster within this boundary.
\\

The column density of molecular hydrogen can be calculated from the velocity integrated main-beam temperature ($T_{mb}=\frac{T_A}{\eta_{mb}}$):
\begin{equation}
{\rm N_{H_2}} = {\rm \frac{H_2}{CO}} \frac{8\pi\nu^3}{c^3 A_{ul}}\frac{1}{e^{h\nu_0/kT_{ex}}-1}\frac{1}{  J(T_{ex})} \frac{Q(T)}{g_u} e^{E_u/kT_{ex}}\int T_{mb} d\upsilon,
\label{eq1}
\end{equation}
where $A_{ul}$ is Einstein coefficient, Q(T) is the partition function, ${\rm \frac{H_2}{CO}}$ is the abundance ratio of molecular hydrogen and CO, $g_u$ is the statistical weight of the states and $J(T_{ex})$ is defined as 

\begin{equation}
J(T_{ex}) \equiv \frac{h\nu}{k}\frac{1}{e^{h\nu/kT_{ex}}-1}.
\end{equation}
Eq. (\ref{eq1}) can be simplified as:

\begin{equation}
{\rm N_{H_2}} ={\rm \frac{H_2}{CO}} 1.94\times 10^{3} \frac{\nu^2{\rm [GHz]}}{A_{ul}}  \frac{Q(T)}{g_u} e^{E_u/kT_{ex}}\int T_{mb} d\upsilon,
\label{density}
\end{equation}
where $\nu$ is in the unit of GHz, $A_{ul} = A_{32} = 2.6 \times 10^{-6} $, $E_u=33.2 \rm K$.  Here we use ${\rm \frac{H_2}{CO}}  \simeq 10^{5}$ as suggested by \citet{jl05} and T$_{ex}$ is assumed to be 20 K \citep{adam09}.  The peak intensity was found at $\alpha=05^h39^m12^s$ $\delta=+35^{\circ}45'46''$ with $T_A = 107$K, and derived column density is N$_{\rm {H_2}} \sim1.1\times10^{22}$ cm$^{-2}$.  The  total mass of the cloud was estimated to be $\sim1000 M_\odot$.

We also estimated the column density using the relation derived by \citet{frerking82} for dust extinction from  the average extinction value, A$_V \sim 13.0$ magnitude (see \S 3.2), of the S233IR cloud,
\begin{equation}
{\rm N_{H_2}} = {\rm A_V} \times 0.94 \times 10^{21}  \rm cm^{-2}.
\end{equation}
The derived column density of S233IR is $1.2\times10^{22}$ cm$^{-2}$, which gives a total gas mass 1100 $M_\odot$, within 30 K km s$^{-1}$ contour (1.5 pc) in diameter.  We derived similar values in two methods mentioned above, but adopted the result from the dust extinction since the infrared result may better correlated with dust emission.


\subsection{Stellar photometry and extinction correction}\label{photometry}


S233IR is a complicated region, which contains stellar clusters embedded,  extended nebulous emissions and shocked outflow gas, as shown in Fig.~\ref{aro}(b).  Properties of outflow gas in $H_2$ band is discussed by \cite{adam09} and the extended emissions will be discussed in detail in the forthcoming paper. Here we focus on the properties of the embedded stellar sources.  The identified point sources  within the cloud boundary were categorized into 3 groups, the SW cluster, the NE cluster  and the  ``distributed sources'' associated with the molecular cloud S233IR except of the NE (Fig.~\ref{aro}(c)) and SW clusters (Fig.~\ref{aro}(d)).  The centers of  NE and SW are at $\alpha=05^h39^m12^s$ $\delta=+35^{\circ}45'59"$ and  $\alpha=05^h39^m09^s$ $\delta=+35^{\circ}45'10"$, respectively.  The radii of both clusters are set to be about 30$\arcsec$ based on the morphology and distribution of extended nebulae.

\begin{figure}[h]
\centering
\plotone{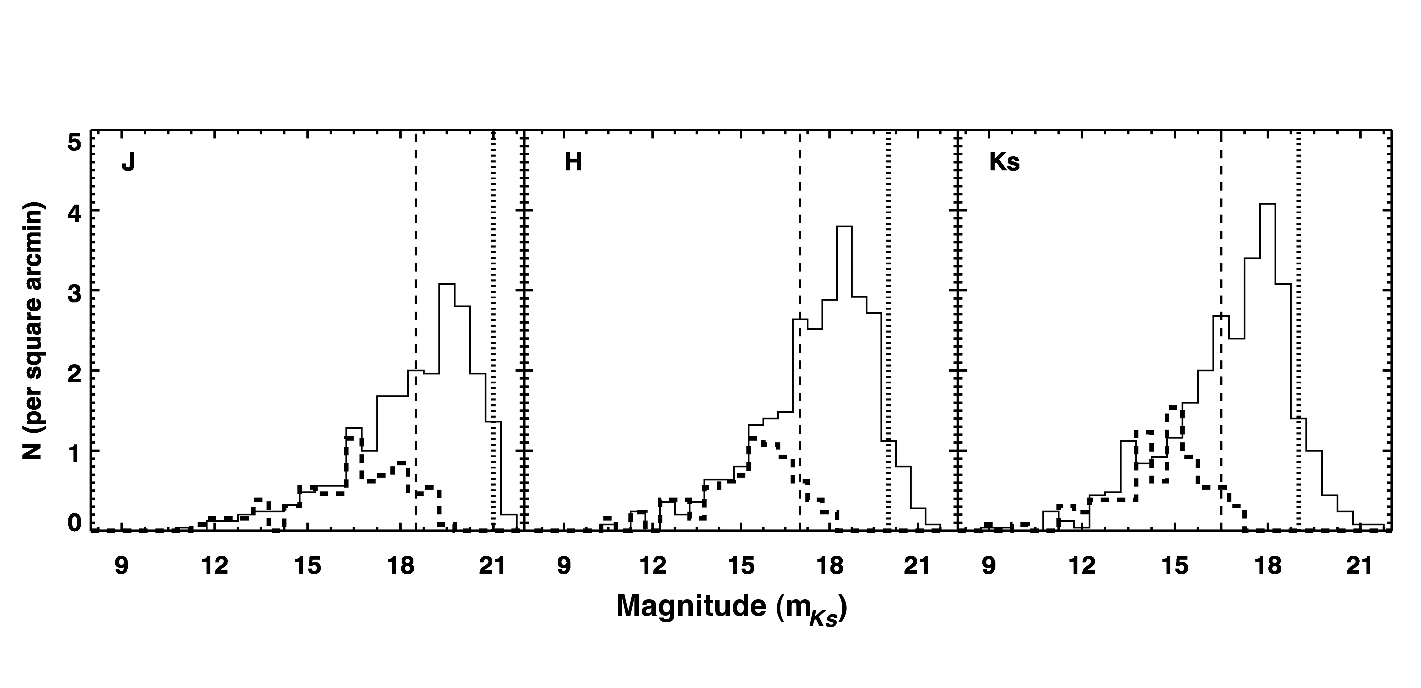}
\caption{Histograms of star counts in different filters.  The histograms with solid and dashed lines are our data and data by \cite{porras00}, respectively.  Vertical lines represent the 90\% completeness limits of this work (dotted line) and of \cite{porras00} (dashed line).}
\label{starcount} 
\end{figure}

Fig.~\ref{starcount} shows the luminosity histogram of the identified point sources (solid line) in different filter bands ($J$, $H$ and $K_S$) with 0.5 magnitude bin.    Comparing with previous data by \citet{porras00}, our observations were made with deeper limiting magnitudes ($\sim$ 2 magnitude) and better image resolution ($\sim$ 3 time better), which results in much large number of detected sources.  For example, a bright source (IR 93) in the NE cluster, which was considered as one point source in \citet{porras00} data, was resolved into several fainter sources in our data.  The two histograms are almost identical for the sources brighter than $\sim 15$ magnitude.

\begin{figure}[h]
\centering
\plotone{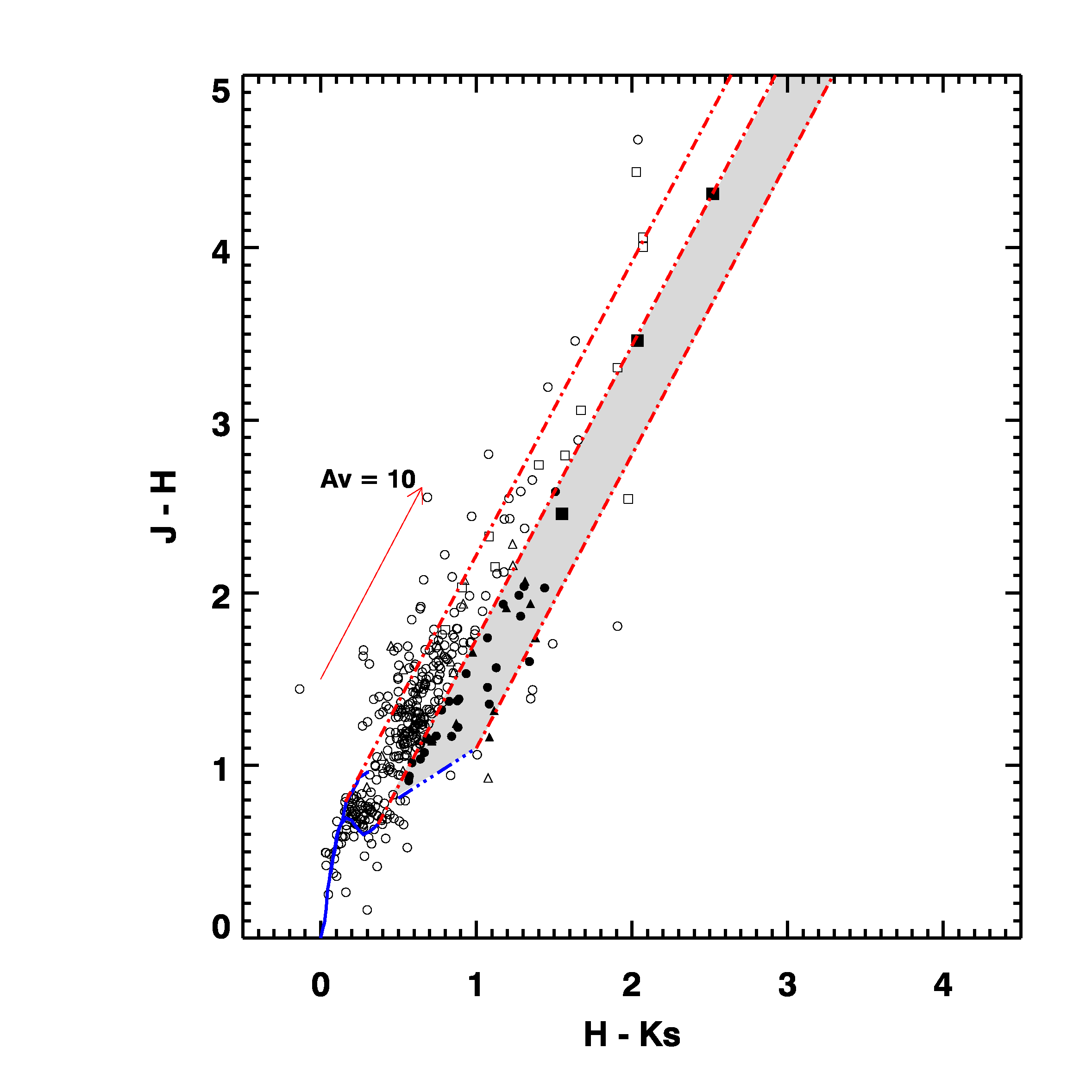}
\caption{The $H-K_S$ vs. $J-H$ diagram of the identified point sources in the S233IR region.   Sources found in the SW,  NE cluster regions, and the ``distributed sources'' are marked with squares, diamonds, and circles, respectively.   CTTS are shown with filled symbols in the gray-shaded region.  The blue dashed line is the CTTS locus and the solid blue lines the track of main sequence stars \citep{bb88, meyer96}.  Red dashed lines indicated the direction of extinction, and the red arrow is for $A_V=10$.  
}
\label{hr}
\end{figure}

\begin{figure}[h]
\centering
\resizebox{12cm}{!}{\rotatebox{0.}{\includegraphics{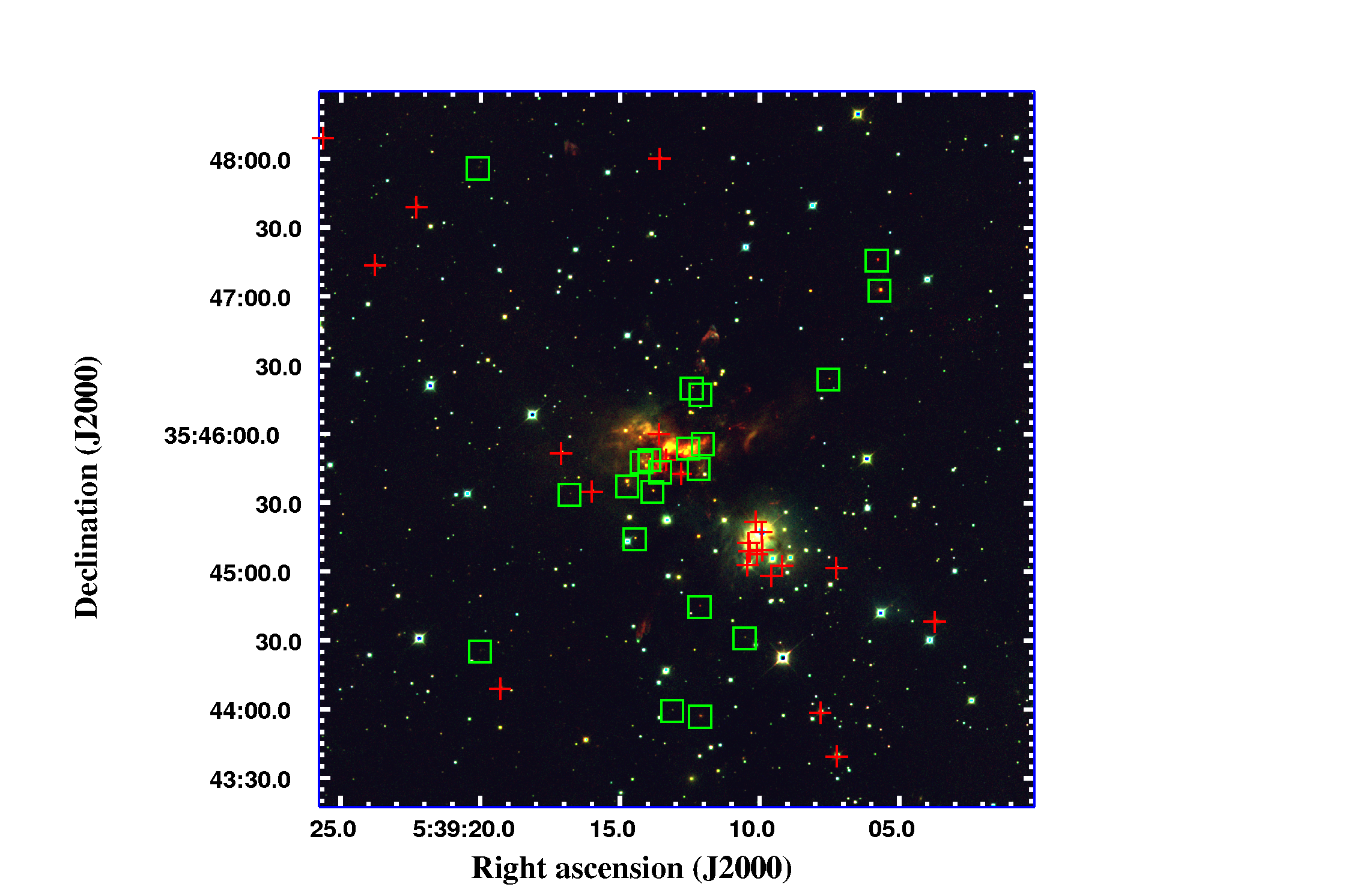}} }
\caption{Location of the sources identified as CTTSs (red crosses) and infrared-excess sources (stars detected in $H$ and $K_S$, green boxes).  The background color image is the same as Fig.~\ref{aro}(b).
}
\label{ysomark}
\end{figure}

Fig.~\ref{hr} is the color-color diagram ($H-K_S$ vs. $J-H$) for the identified sources in the S233IR cloud.  The stars detected in the SW and NE cluster regions are marked with squares and triangles, respectively, and the other sources (``distributed sources'') are marked with circles.  Filled symbols are for the sources within the classical T-Tau stars (CTTSs) range.  The fraction of IR excess (non-$J$ band detection) stars, which include the stars in the CTTS region or not detected in the $J$ band, are 35\% (12/34) and 36\% (13/26) for the SW and NE clusters, respectively.   The CTTS and  infrared-excess ($H-K_S > 1.5$) sources are identified in Fig.~\ref{ysomark}, with red crosses and green boxes, respectively.   Some very red objects located at outskirts of this molecular seem to be  isolated star-forming sources.  The nature of these sources will be discussed in the second paper.  

We identified about 800 sources in total within the cloud boundary.  It was expected that a significant number of background sources exist toward the target cloud in the observed bands.  It is an important but very difficult job to separate the member sources from the background.   Without proper motion data for the cluster, statistical subtraction is the only way to determine the member stars.  To subtract the background stars,  we chose a region outside the molecular boundary, centered at $\alpha=05^h38^m40.1^s$  $\delta=+35^{\circ}42' 07.7"$, as a reference field with {\it no} extinction.  The reference field was chosen in the area well outside the cloud boundary based on radio observations.   By assuming that there exists the {\it same} background stars in the target cloud (the S233IR cloud), we removed expected background sources statistically, following the procedure summarized by \cite{jose08}: select a star on the color-magnitude diagram (CMD) of the reference field,  eliminate the star with same color and magnitude on the CMD of the target field within the observed uncertainties ($\Delta (H-K_S) \leq$ 0.08 and $\Delta K_S \leq$ 0.07). By repeating this procedure for all sources in CMD, we subtract background sources in the target field.  There were 254 sources cleaned in a 5$\arcmin$ radius region.  This leads to a number of $\sim2$ background sources in 30$\arcsec$ radius region.  In other words, 92\% and 94\% of the sources are associated with SW and NE cluster, respective.

\begin{figure}[h]
\centering
\resizebox{12cm}{!}{\rotatebox{0.}{\includegraphics{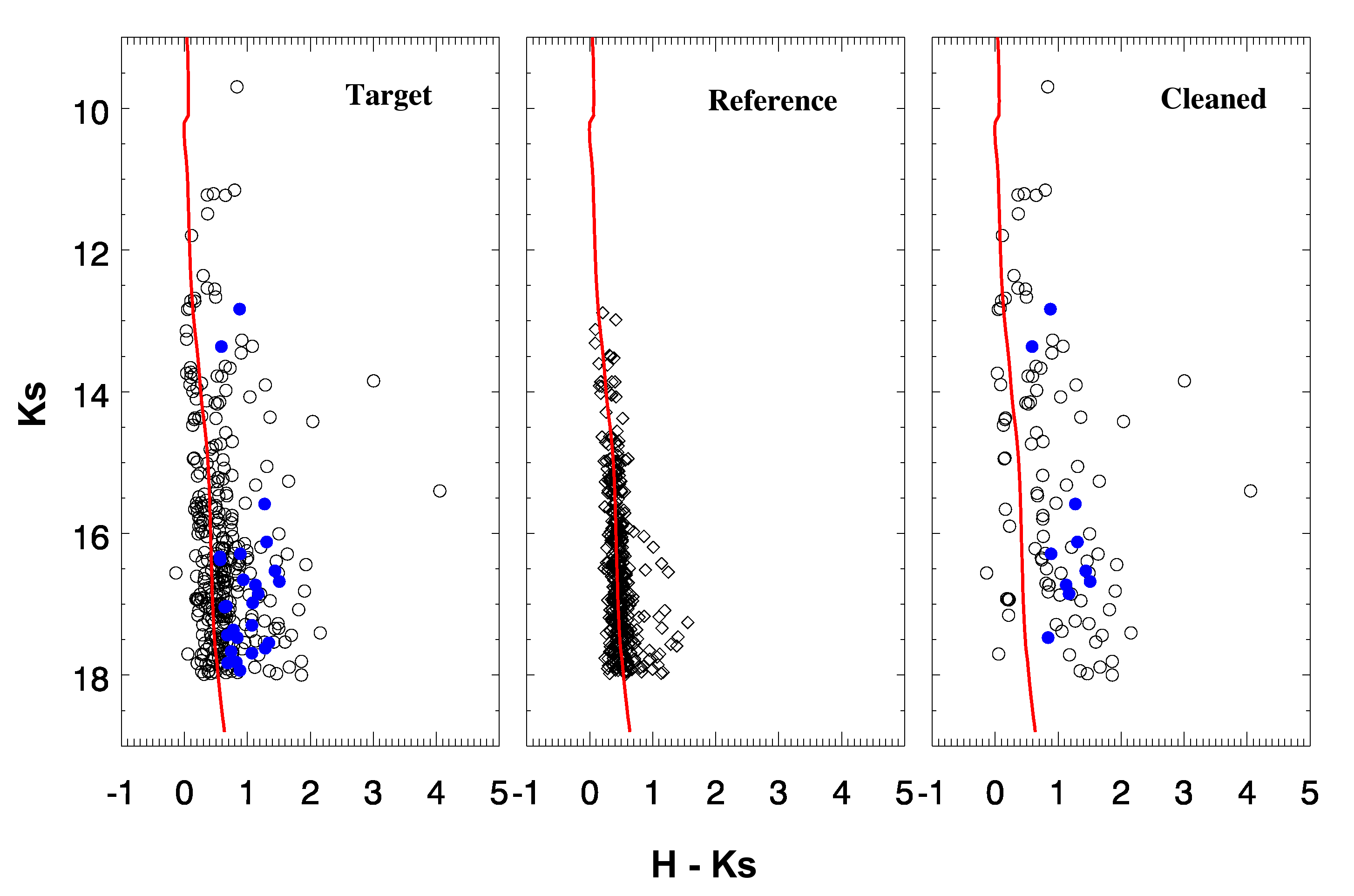}} }
\caption{The color-magnitude ($K_S$ vs. $H-K_S$) diagrams.   
Source symbols are the same as in Fig.~\ref{hr}.  
Red lines represent the age-averaged isochrone between 1.0 to 10.0 Myrs from \cite{dm94}. }
\label{cmd}
\end{figure}

Fig.~\ref{cmd} shows the color-magnitude diagram ($K_S$ vs. $H-K_S$) for the sources in the S233IR cloud, the reference field, and the background cleaned ``member sources'', respectively.  The red lines in the panels represent the age-averaged isochrone between 1.0 to 10.0 Myrs from \cite{dm94}.    Although there should exist various uncertainties in this process, we think that this method is the most efficient way to determine the member sources at least within the observational uncertainties.   Additionally, stars on the left side of the isochrone in Fig.~\ref{cmd} were removed as foreground stars.  The final result, after the background elimination, is shown in the right panel of Fig. \ref{cmd}. 

With these ``member'' sources, we made extinction correction for pre-main sequence stars (PMSs) and CTTSs with two different ways, respectively, using two diagrams (Fig.~\ref{hr} and Fig.~\ref{cmd}).  Extinction corrections for PMSs were made by projecting them back to the mean isochrone along the reddening vector on Fig.~\ref{cmd}, $(H-K_S)^{observed}=(H-K_S)^{true}+0.063 \times A_V$, by assuming a normal reddening law \citep{rl85}.  The error of dereddened $K_S$ magnitudes is expected $\leq 0.2$ from the mean isochrone we used \citep{massi06}. The extinction can be derived by comparing observed and intrinsic color of each star. 
On the other hand, extinctions of the CTTSs (filled symbols in the figures) were corrected by projecting them back to the CTTS locus on Fig.~\ref{hr}.   The averaged visual extinction ($A_V$) were found to be 9.8$\pm$5.2 (the SW cluster, 36 stars), 28.9$\pm$10.4 (the NE cluster, 25 stars), and 13.0$\pm$10.4 (``distributed stars'', 124 stars).  These extinction are larger than the previous values by a factor of about $1.3-2$ (PCS), which result from deeper observations and also from more reliable analysis by using both diagrams (Figs.~\ref{hr}, and \ref{cmd}) in deriving extinction amounts.

\subsection{KLF modeling}\label{klfmodel}
For decades, the most reliable method in optical astronomy used to determine the age of a cluster is using the HR diagram (HRD), comparing the positions of member stars with theoretical evolutionary tracks on the HRD.  For embedded young cluster, this method is not validated because most of the members are not in the main sequence stage and can be observed only in longer wavelengths than optical.  The K-band luminosity function (KLF) is a simple tool to study the properties and estimate the age of an embedded cluster \citep{lada03,yasui06}.  Definition of the KLF can be expressed by the following
equation:
\begin{equation}
\frac{dN}{dm_k} = \frac{dN}{d\textrm{log} M_*} \times \frac{d\textrm{log}
M_*}{dm_k}, 
\end{equation}
where $m_k$ is the K-band luminosity and $M_*$ is the stellar mass \citep{lada03}.  The first term in the right hand side is the underlying stellar mass function and the second term is the mass-luminosity relation (MLR).  It is noticed that the KLF of clusters peak at different magnitude, depending on  the difference between their ages and star formation history \citep{muench00}.  

Simple Monte Carlo simulations were carried out to construct the model KLFs.  The simulation was done in three step.  The first step is to assume an initial mass function (IMF).  Two IMFs were used in our simulation, they are Trapezium IMF \citep{muench00} and the IMF from \citet[][hereafter MS79]{ms79}.  Then, we convert the mass function to the luminosity function using a mass-to-luminosity relation (MLR) from the isochrones of the PMS models\citep{dm94,dm97,dm98,siess00}.   The stellar luminosities were finally converted to the K-band luminosity, $m_K$ with bolometric correction \citep{flower96} and stellar intrinsic color correction \citep{bb88}.   By repeating this procedure with different age inputs, we fit model KLFs to observed KLFs and estimate the ages of clusters.  In Fig.~\ref{clusterage}, we show a result derived for embedded clusters in S233IR, tp be discussed in the following section.


\subsection{Age estimates from KLFs and derived IMFs of the embedded clusters} \label{klf-imf}
\begin{figure}[h]
\centering
\resizebox{14cm}{!}{\rotatebox{0.}{\includegraphics{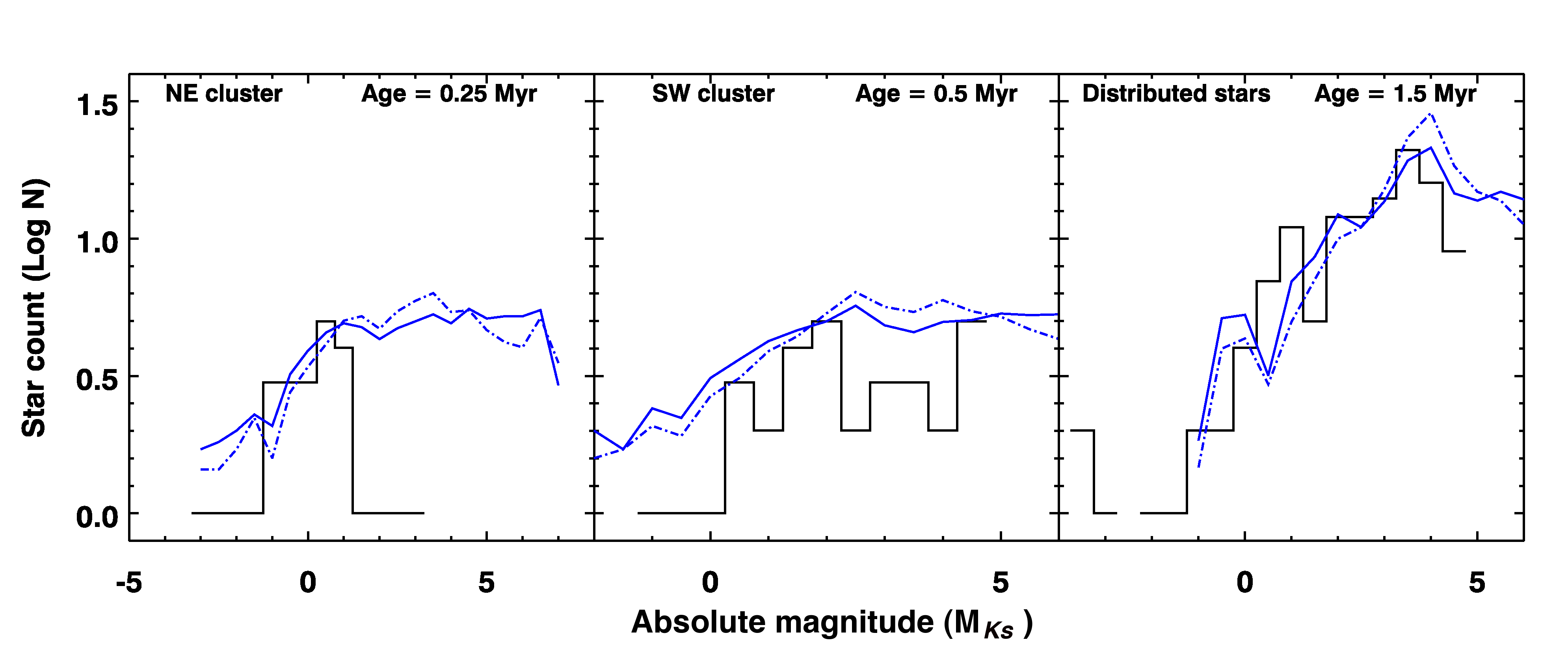}} }
\caption{Histograms of K-band luminosity function in the SW, NE, and the distributed stars as indicated in the panels.  The solid and dash-dotted lines represents the model KLFs based on the IMF of \citet{ms79} and \citet{muench02}, respectively.}
\label{clusterage}
\end{figure}

Ages of the clusters embedded in the S233IR cloud were estimated by comparing the observed and model KLFs \citep{yasui06}.  Using the method mentioned  in \S \ref{klfmodel}, we constructed model KLFs for ages from 0.07 to 50 Myrs.   By comparing the model and observed KLFs (Fig.~\ref{clusterage}), we derived the ages of $\sim0.5\pm0.1$, $\sim0.25\pm0.1$, and $\sim1.5\pm0.3$ Myrs for the SW, NE, and the distributed stars, respectively.  Although the MLR for the age younger than 1 Myr is uncertain between different PMS models \citep{baraffe02,yasui06}, this method allows us estimating cluster ages quantitatively.  The age difference of these stellar groups was first noticed by \citet{porras00} based on the $J$-band data analysis.  It is understood that there are uncertainties in estimating cluster ages because we detect 25 and 36 IR sources in NE and SW cluster, respectively.  However,  the model KLFs are not  sensitive to the initial mass functions (IMFs) used.  Therefore, the current results do suggest that relative age differences exist among these clusters roughly with the amounts mentioned here.  Using the estimated ages of the clusters, the luminosity function was converted to the mass with the model isochrones \citep{dm94,dm97,dm98,siess00} by applying the bolometric correction \citep{flower96} to the expected stellar intrinsic colors \citep{bb88}.  The total masses of $\gtrsim$46, $\gtrsim$30, and $\sim$110 $M_\odot$ were derived for the NE, SW, and the distributed stars, respectively.

With the determined age, the IMFs of the embedded clusters can be calculated using the mass-luminosity relation and its slope $\Gamma$, which is defined as $dN / d log(m) = M_*^{\Gamma}$ \citep[cf.][]{muench02}, to constrain the cluster properties.  The size of the IMF bin was set to be 0.5 M$_\odot$, which is larger than the uncertainty of mass estimation propagated from luminosity error.   We derived $\Gamma \sim -0.42\pm0.11$, $\sim -0.10\pm0.08$, and $\sim -0.46\pm0.07$ for the SW , NE, and  distributed stars, respectively.  Although the derived $\Gamma$ values for the NE and SW clusters are uncertain due to small numbers (25 source for NE and 36 sources for SW),  it still indicates that the $\Gamma$ values in this region are  much flatter than the Salpeter slope.  This may suggest that more massive stars have efficiently formed in the S233IR region compared to other star forming regions in our Galaxy, such as the sources listed by \cite{muench02,fig02,fig05,leistra05,leistra06,jose08,hara08,paney08}, etc., who have derived slopes $\Gamma\sim -1$ in young embedded stellar clusters.

\subsection{Star formation efficiency} \label{sfe}
The star formation efficiency (SFE) is defined as the ratio of the total stellar mass to the total stellar and gas mass \citep[cf.][]{wl83}.  Typically, the SFE ranges from $\sim$10\% to 30\% \citep[][see Table 2 and reference there in]{lada03}.  SFEs have been found low in young or low mass star formation regions (eg. Serpens, Rho Oph and NGC~1333) but increased to about 30\% in more evolved massive star-forming regions \citep{lada03}.  We have determined the total gas mass of S233IR, $\sim 1100~ M_\odot$, and the embedded cluster mass, $\sim 180 ~M_\odot$ (\S.~\ref{klf-imf}).  The SFE is $\sim$17\% for the S233IR region.     Our analysis covers the whole S233IR region and the member stars were determined more carefully than previous studies for this region \cite[for example,][]{beuther02a, beuther02b,mz04}.  

\begin{figure}[h]
\centering
\resizebox{12cm}{!}{\rotatebox{0.}{\includegraphics{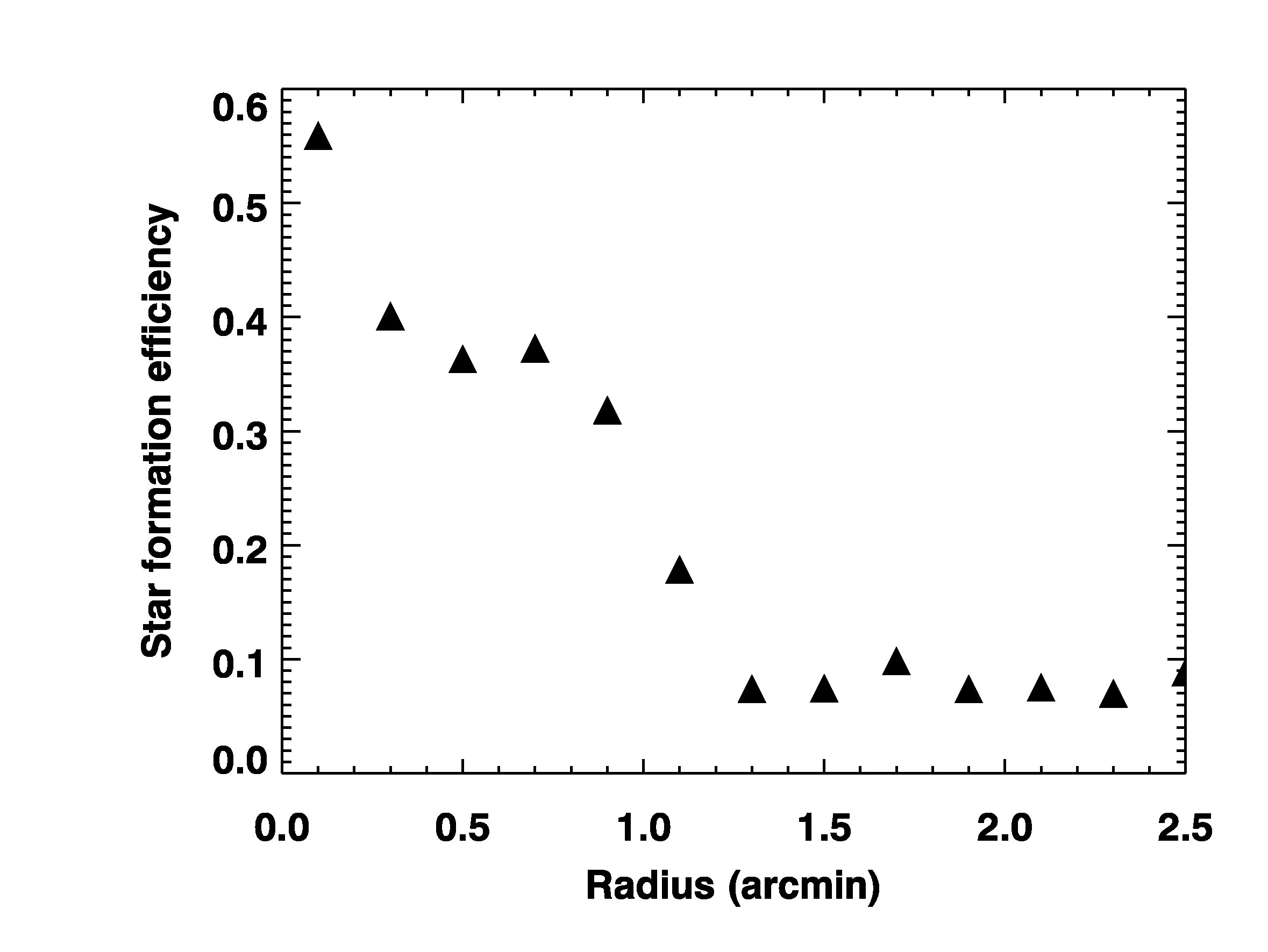}} }
\caption{The radial dependence of star formation efficiency in S233IR region. The SFEs were derived by comparing the stellar mass to the total gas mass along the ring, centered at $\alpha=05^h39^m11.8^s$  $\delta=+35^{\circ}45' 51.1"$, with annuli of width $\Delta r=0.2\arcmin$.  }
\label{sfeprofile}
\end{figure}

It should be noticed that the derived SFE may represent the lower limit, since significant number of sources have not been identified because of the extended emission features and high extinction environments in the central parts of the embedded clusters, this caveat also applies to $\Gamma$.  The latest star-forming cores are found in the NE cluster. Young massive stars are still deeply embedded in the dense molecular cores \citep{beuther07,leurini07} and associated with maser emissions and energetic shocked H$_2$ gas \citep{menten91,porras00, adam09}.   Because of the age differences among stellar groups, it is possible that star formation has not resulted from the overall collapse of the  S233IR molecular could at the same time.  If stars form continuously, the single SFE value for the whole cloud may not represent a measure of the final gas-to-stars efficiency.  We therefore investigated SFEs at different radii in the S233IR cloud.  Since the total gas mass in local areas is difficult to separate from the whole cloud, the SFEs were derived by comparing the stellar mass and the total gas mass along the ring, centered at the cloud center, of the projected width of about $\Delta r = 0.2\arcmin$ (projected distance of $\sim$0.1 pc), as shown in Fig.~\ref{sfeprofile}.  The center is at $\alpha=05^h39^m11.8^s$  $\delta=+35^{\circ}45' 51.1"$, which corresponds to the peak of the integrated CO emission.  As expected, the SFE changes with the location in the cloud.    High SFE values up to $\gtrsim40$\% are found in the central region ($r\leq1\arcmin$).  This radius covers the region of the NE and SW clusters and both of them represent later stages of star formation.  The high value may suggest that there are local enhancements in massive star formation, possibly triggered by nearby star formations within the S233IR cloud.

\section{Discussion}\label{discuss}

Based on the radio and extinction map \citep{heyer96,ry08}, the S233IR molecular cloud is a part of the much larger Auriga GMC complex, which is  associated with several large HII regions and active star-forming regions.  The large scale ($\sim$ a few degrees) morphology of this GMC complex looks very clumpy and is composed of several HII regions, such as Sh~2-231, 232, 233 and 235 \citep{ry08}.   The CO emission shows a filamentary and arc-like structure associated with three HII regions.   At the northeast direction, an HII region,  Sh~2-231, is $\sim10\arcmin$ away from S233IR  \citep{mz04}.  In addition, there are many newly formed OB stars, such as those in Aur OB1, associated with this GMC complex \citep[][Figure 1]{ry08}.  The S233IR clump appears to be the latest of star formation events.  

Three groups of stars were identified in S233IR region, SW, NE and the distributed stars.  The background contamination in distributed stars were eliminated by a statistical method.  The age differences among these subgroups were first noticed by \citet{porras00} using the $J$-band analysis.  Based on better data quality of our work, we derived ages $\sim$0.5 Myrs for SW, $\sim$0.3 Myrs for NE, and $\sim$1.5 Myrs for the distributed stars.   The distributed stars could be the first generation of stars formed in this cloud.  The SW and NE clusters, with the relative distance of about 0.5 pc at a 1.8 kpc distance, are later generations.   The age differences between stellar groups suggests sequential star formation, which has also been found in other star-forming regions, such as NGC~7538 \citep{ojha04}, Sh~2-247 \citep{puga08}, Sh~2-157 \citep{chen09} and NGC~3576\citep{purcell09}.  It is interesting that the S233IR cloud, a typical star-forming cloud in size ($r\sim$1.5 pc) and mass ($\sim1100$ M$_\odot$), contains sub-clusters formed in different epochs and star formation still continues without disrupting its natal molecular cloud.  Stars appear to be forming continuously in the youngest NE cluster.   For the distributed stars, the derived IMF slope,  $\Gamma\sim -0.5$, is flatter than in other  star-forming regions in our Galaxy,   which indicates that more massive stars have been formed preferentially in this cloud than in other typical star-forming regions.   The $\Gamma$ value of the NE cluster, the latest star-forming sub-core in this cloud, is even flatter than the distributed stars,  suggesting that the probability of massive star formation is higher at this region.   This result is consistent with the radio observation toward the center of NE cluster \citep{beuther02a}.   Eventually the cluster will disperse the natal molecular cloud and appear as visible stars.  Once the massive stars are formed, the radiation from massive star will clear the molecular could and stop the star formation process in the core region.


The enhancement of  the local SFE is also found in the central dense core as we derived in \S\ref{sfe}, although  we have to take various uncertainties into consideration, for example, the uncertain total H$_2$ density, the cloud boundary, member star determination, undetected sources especially near the central part of the sub-clusters showing saturated infrared emission, etc.  The local high SFE ( $\gtrsim$40\%) suggests that the later generation of stars can be formed very efficiently in dense cores in the same molecular cloud.    In the outskirt of the cloud, where the first stellar generation is located, the SFE is consistent with a typical molecular cloud.    The trigged star formation may have enhanced the formation of massive stars, which  results in a high SFE and flatter IMF slope in the central region.  It is known that the NE cluster is the latest star forming core in S233IR, which is associated with the powerful energetic features, such as shocked H$_2$ gas flows \citep{adam09}.  These energetic activities are thought to be a part of massive star formation process.   The S233IR region provides an interesting example of active and energetic massive star formation, which could be further studied with next generation instruments.

\section{Summary}\label{summary}

We revisited the S233IR (Sh~2-233IR) region with wider field of view and more sensitive near-infrared data than previous studies.  There exist three distinguishable sub-clusters in this region: the NE and SW clusters, and the distributed stars.  Using the color-color ($H-K_S$ vs. $J-H$) and the color-magnitude ($K_S$ vs. $H-K_S$) diagrams, we subtracted background stars statistically and identified 25, 33, and 151 stellar sources for the NE cluster, the SW cluster, and the distributed stars, respectively.  We derived ages $\sim$0.3, $\sim$0.5, and $\sim$1.5 Myrs, and mass 45, 30, and 107 $M_\odot$, for  SW, NE and the distributed stars, respectively.  The flatter IMF slope ($\Gamma\gtrsim-0.5$) for subclusters suggests that more massive stars were preferentially formed in S233IR.  The first stellar generation is the distributed stars located in the GMC complex in which S233IR is embedded.  Afterward,  the sequential star formation in this cloud may have been triggered before its natal molecular cloud is disturbed by previous star formation events.   The averaged SFE of this region is $\sim$17\%, and $\gtrsim$40\% at the center of S233IR, the NE and SW clusters.  This result suggests that stars form very efficiently in local dense cores within the cloud center.

\vspace{0.5cm}

{\bf Acknowledgements.} 
We thank the staff of the Canada-France-Hawaii Telescope (CFHT) and Arizona Radio Observatory (ARO) for their very competent assist in obtaining the data used in this work.  Access to the CFHT was made possible by the Ministry of Education and the National Science Council of Taiwan as part of the Cosmology and Particle Astrophysics (CosPA) initiative.  This research has made use of  the VizieR catalogue access tool, CDS, Strasbourg, France. YCM acknowledges support by the Korea Science and Engineering Foundation (KOSEF) grant funded by the Korea government (MEST) (No.\ 2009-0083999), and by the Korea Research Foundation (KRF) grant (KRF-2008-313-C00376).


\vspace{0.5cm}

\begin{thebibliography}{}
\bibitem[Andr{\'e}(1994)]{andre94} Andr{\'e}, P.\ 1994, In \textit{The 
Cold Universe}, 179, Montmerle, T., Lada, C. J., Mirabel, I. F. and Tr\^{a}n
Tranh V\^{a}n, J. (eds.) , Editions Fronti\`{e}res, Gif-sur-Yvtte, p. 179

\bibitem[Baraffe et 
al.(2002)]{baraffe02} Baraffe, I., Chabrier, G., Allard, F., \& Hauschildt, P.~H.\ 2002, \aap, 382, 563 

\bibitem[Bessell \& Brett(1988)]{bb88} Bessell, M.~S., \& Brett, J.~M.\ 1988,
\pasp, 100, 1134 


\bibitem[Beuther et al.(2007)]{beuther07} Beuther, H., Leurini, S., Schilke, P.,
Wyrowski, F., Menten, K.~M. \& Zhang, Q.\ 2007, \aap, 466, 1065 

\bibitem[Beuther et al.(2002a)]{beuther02a} Beuther, H., Schilke, P., Gueth, F.,
McCaughrean, M., Andersen, M., Sridharan, T.~K. \& Menten, K.~M.\ 2002, \aap,
387, 931 

\bibitem[Beuther et al.(2002b)]{beuther02b} Beuther, H., Walsh, A., Schilke, P.,
Sridharan, T.~K., Menten, K.~M., \& Wyrowski, F.\ 2002, \aap, 390, 289

\bibitem[Bonnell 
\& Bate(2005)]{bb05} Bonnell, I.~A., \& Bate, M.~R.\ 2005, \mnras, 362, 915 

\bibitem[Casoli et al.(1986)]{casoli86} Casoli, F., Combes, F.,
Dupraz, C., Gerin, M., \& Boulanger, F.\ 1986, \aap, 169, 281 

\bibitem[Cesaroni et al.(1999)]{cesaroni99} Cesaroni, R., Felli, M., \&
Walmsley, C.~M.\ 1999, \aaps, 136, 333 

\bibitem[Chen et al.(2009)]{chen09} Chen, Y., Yao, Y., Yang, 
J., Zeng, Q., \& Sato, S.\ 2009, \apj, 693, 430 

\bibitem[D'Antona \& Mazzitelli(1994)]{dm94} D'Antona, F., \& Mazzitelli, I.\
1994, \apjs, 90, 467 

\bibitem[D'Antona \& Mazzitelli(1997)]{dm97} D'Antona, F., \& Mazzitelli, I.\
1997, Memorie della Societa Astronomica Italiana, 68, 807 


\bibitem[D'Antona \& Mazzitelli(1998)]{dm98} D'Antona, F., \& Mazzitelli, I.\
1998, in ASP Conf. Ser. 134, Brown Dwarfs and Extrasolar Planets, ed. R. Rebolo,
E. L. Martin \& M. R. Zapatero Osorio (San Francisco: ASP), 442 

\bibitem[Ginsburg et al.(2009)]{adam09} Ginsburg, A., Bally, J. \& Yan, C. -H.
2009, accepted by \apj

\bibitem[Figuer{\^e}do et al.(2005)]{fig05} Figuer{\^e}do,
E., Blum, R.~D., Damineli, A., \& Conti, P.~S.\ 2005, \aj, 129, 1523 


\bibitem[Figuer{\^e}do et al.(2002)]{fig02} Figuer{\^e}do,
E., Blum, R.~D., Damineli, A., \& Conti, P.~S.\ 2002, \aj, 124, 2739 

\bibitem[Frerking et al.(1982)]{frerking82} Frerking, M.~A., 
Langer, W.~D., \& Wilson, R.~W.\ 1982, \apj, 262, 590 

\bibitem[Flower(1996)]{flower96} Flower, P.~J.\ 1996, \apj, 469, 
355
 
\bibitem[Harayama et al.(2008)]{hara08} Harayama, Y., 
Eisenhauer, F., \& Martins, F.\ 2008, \apj, 675, 1319 


\bibitem[Henning et al.(1992)]{henning92} Henning, T., Cesaroni, R., Walmsley,
M. \& Pfau, W.\ 

\bibitem[Heyer et al.(1996)]{heyer96} Heyer, M.~H., Carpenter, 
J.~M., \& Ladd, E.~F.\ 1996, \apj, 463, 630 

\bibitem[Hunter \& Massey(1990)]{hunter90} Hunter, D.~A. \& 
Massey, P.\ 1990, \aj, 99, 846 

\bibitem[Hunter(1992)]{hunter92} Hunter, D.~A.\ 1992, \apjs, 79, 469 

\bibitem[Hodapp(1994)]{hodapp94} Hodapp, K.-W.\ 1994, \apjs, 94, 
615

\bibitem[Jiang et al.(2000)]{jiang00} Jiang, Z., Wang, M., 
\& Yang, J.\ 2000, Acta Astronomica Sinica, 41, 28 

\bibitem[Jiang et al.(2001)]{jiang01} Jiang, Z., Yao, Y., Yang, J., Ishii, M.,
Nagata, T., Nakaya, H. \& Sato, S.\ 2001, \aj, 122, 313 

\bibitem[Jose et al.(2008)]{jose08} Jose, J., et al.\ 2008,
\mnras, 384, 1675 

\bibitem[Lada 
\& Lada(2003)]{lada03} Lada, C.~J., \& Lada, E.~A.\ 2003, \araa, 41, 57 


\bibitem[Larionov et al.(1999)]{lari99} Larionov, G.~M., Val'tts, I.~E.,
Winnberg, A., Johansson, L.~E.~B., Booth, R.~S. \& Golubev, V.~V.\ 1999, \aaps,
139, 257 

\bibitem[Leistra et al.(2005)]{leistra05} Leistra, A., Cotera,
A.~S., Liebert, J., \& Burton, M.\ 2005, \aj, 130, 1719 



\bibitem[Leistra et al.(2006)]{leistra06} Leistra, A., Cotera,
A.~S., \& Liebert, J.\ 2006, \aj, 131, 2571 


\bibitem[Leurini et al.(2007)]{leurini07} Leurini, S., Beuther, 
H., Schilke, P., Wyrowski, F., Zhang, Q., \& Menten, K.~M.\ 2007, \aap, 
475, 925 

\bibitem[Lequeux(2005)]{jl05} Lequeux, J, 2005, {\it The Interstellar Medium}, Springer-Verlag, Berlin 

\bibitem[Longmore et al.(2006)]{longmore06} Longmore, S.~N., 
Burton, M.~G., Minier, V. \& Walsh, A.~J.\ 2006, \mnras, 369, 1196 


\bibitem[Mao \& Zeng(2004)]{mz04} Mao, R.-Q. \& Zeng, Q.\ 
2004, Chinese Journal of Astronomy and Astrophysics, 4, 440 

\bibitem[Massi et 
al.(2006)]{massi06} Massi, F., Testi, L., \& Vanzi, L.\ 2006, \aap,
448, 1007 

\bibitem[Menten(1991)]{menten91} Menten, K.~M.\ 1991, \apjl, 
380, L75 

\bibitem[Meyer(1996)]{meyer96} Meyer, M.~R.\ 1996, 
Ph.D.~Thesis, University of Massachusetts 


\bibitem[Minier et al.(2005)]{minier05} Minier, V., Burton, M.~G., Hill, T.,
Pestalozzi, M.~R., Purcell, C.~R., Garay, G., Walsh, A.~J. \& Longmore, S.\
2005, \aap, 429, 945

\bibitem[Miller \& Scalo(1979)]{ms79} Miller, G.~E., \& Scalo,
J.~M.\ 1979, \apjs, 41, 513 

\bibitem[Motoyama \& Yoshida(2003)]{mot03}Motoyama, K. \& Yoshida, T. 2003, \mnras, 344, 461

\bibitem[Muench et al.(2000)]{muench00} Muench, A.~A., Lada, 
E.~A., \& Lada, C.~J.\ 2000, \apj, 533, 358

\bibitem[Muench et al.(2002)]{muench02} Muench, A.~A., Lada, 
E.~A., Lada, C.~J., \& Alves, J.\ 2002, \apj, 573, 366 

\bibitem[Ojha et al.(2004)]{ojha04} Ojha, D.~K., et al.\ 2004, 
\apj, 616, 1042 

\bibitem[Pandey et al.(2008)]{paney08} Pandey, A.~K., Sharma,
S., Ogura, K., Ojha, D.~K., Chen, W.~P., Bhatt, B.~C., 
\& Ghosh, S.~K.\ 2008, \mnras, 383, 1241 


\bibitem[Pascal et al.(2004)]{pascal04} Pascal Puget, et al., 2004,  Proc.
SPIE. 5492, 978

\bibitem[Porras et al.(2000)]{porras00} Porras, A., Cruz-Gonz{\'a}lez, I. \&
Salas, L.\ 2000, \aap, 361, 660

\bibitem[Puga et al.(2008)]{puga08} Puga, E., et al.\ 2008, 
Massive Star Formation: Observations Confront Theory, 387, 331 

\bibitem[Purcell et 
al.(2009)]{purcell09} Purcell, C.~R., et al.\ 2009, \aap, 504, 139 

\bibitem[Reipurth \& Yan(2008)]{ry08} Reipurth, B \& Yan, C.-H., 2008, in ASP Conf. Ser. 402, \textit{Handbook of Star Forming Regions: Volume I the Northern Sky}, eds. B. Reipurth

\bibitem[Rieke \& Lebofsky(1985)]{rl85} Rieke, G.~H., \& Lebofsky, M.~J.\ 1985,
\apj, 288, 618 


\bibitem[Shepherd 
\& Churchwell(1996)]{sc96} Shepherd, D.~S., \& Churchwell, E.\ 1996, \apj, 472, 225 

\bibitem[Shepherd 
\& Watson(2002)]{sw02} Shepherd, D.~S., \& Watson, A.~M.\ 2002, \apj, 566, 966 

\bibitem[Siess et 
al.(2000)]{siess00} Siess, L., Dufour, E., \& Forestini, M.\ 2000, \aap, 358, 593 

\bibitem[Snell et al.(1990)]{snell90} Snell, R.~L., Dickman, 
R.~L. \& Huang, Y.-L.\ 1990, \apj, 352, 139 

\bibitem[Sridharan et al.(2002)]{sri02} Sridharan, T.~K., 
Beuther, H., Schilke, P., Menten, K.~M., 
\& Wyrowski, F.\ 2002, \apj, 566, 931 

\bibitem[Tofani et al.(1995)]{tofani95} Tofani, G., Felli, M., 
Taylor, G.~B. \& Hunter, T.~R.\ 1995, \aaps, 112, 299 

\bibitem[Varricatt et al.(2005)]{varricatt05} Varricatt, W.~P., 
Davis, C.~J., \& Adamson, A.~J.\ 2005, \mnras, 359, 2


\bibitem[Wilking \& Lada(1983)]{wl83} Wilking, B.~A., \& Lada, C.~J.\ 1983, \apj, 274, 698

\bibitem[Yao et al.(2000)]{yao00} Yao, Y., Ishii, M., Nagata, T., Nakaya, H. \&
Sato, S.\ 2000, \apj, 542, 392 

\bibitem[Yasui et al.(2006)]{yasui06} Yasui, C., Kobayashi, N., 
Tokunaga, A.~T., Terada, H., \& Saito, M.\ 2006, \apj, 649, 753 

\bibitem[Yan et al.(2010)]{chy09b} Yan, C.-H., Minh, Y.~C.,  \& Wang, S.-Y.\ 2010, in preparation 


\end{thebibliography}
\end{document}